\documentclass[master]{ws-rv9x6} 
\usepackage{ws-rv-har}    
\usepackage{ws-rv-thm}    
\usepackage{subfigure}    
\usepackage{ws-index}     
\usepackage{enumitem}
\usepackage[percent]{overpic}
\usepackage{xcolor}
\makeindex
\newindex{aindx}{adx}{and}{Author Index}       
\renewindex{default}{idx}{ind}{Subject Index}  

\begin{document}		
\newcommand{\ltsima}{$\; \buildrel < \over \sim \;$}
\newcommand{\lsim}{\lower.5ex\hbox{\ltsima}}
\newcommand{\gtsima}{$\; \buildrel > \over \sim \;$}
\newcommand{\gsim}{\lower.5ex\hbox{\gtsima}}
\newcommand{\bra}{\langle}
\newcommand{\ket}{\rangle}
\newcommand{\lprime}{\ell^\prime}
\newcommand{\lpp}{\ell^{\prime\prime}}
\newcommand{\mprime}{m^\prime}
\newcommand{\mpp}{m^{\prime\prime}}
\newcommand{\ci}{\mathrm{i}}
\newcommand{\dd}{\mathrm{d}}
\newcommand{\veck}{\mathbf{k}}
\newcommand{\vecx}{\mathbf{x}}
\newcommand{\vecr}{\mathbf{r}}
\newcommand{\vecv}{\mathbf{\upsilon}}
\newcommand{\vecw}{\mathbf{\omega}}
\newcommand{\vecj}{\mathbf{j}}
\newcommand{\vecq}{\mathbf{q}}
\newcommand{\vecl}{\mathbf{l}}
\newcommand{\vecn}{\mathbf{n}}
\newcommand{\lm}{\ell m}
\newcommand{\that}{\hat{\theta}}
\newcommand{\thatp}{\that^\prime}
\newcommand{\chip}{\chi^\prime}
\newcommand{\hs}{\hspace{1mm}}
\newcommand{\nar}{New Astronomy Reviews}
\def\gsim{~\rlap{$>$}{\lower 1.0ex\hbox{$\sim$}}}
\def\lsim{~\rlap{$<$}{\lower 1.0ex\hbox{$\sim$}}}
\def\Msun {\,\mathrm{M}_\odot}
\def\Jcrit {J_\mathrm{crit}}
\newcommand{\rsun}{R_{\odot}}
\newcommand{\mbh}{M_{\rm BH}}
\newcommand{\Msunyr}{M_\odot~{\rm yr}^{-1}}
\newcommand{\mdot}{\dot{M}_*}
\newcommand{\ledd}{L_{\rm Edd}}
\newcommand{\cmc}{{\rm cm}^{-3}}
\def\gsim{~\rlap{$>$}{\lower 1.0ex\hbox{$\sim$}}}
\def\lsim{~\rlap{$<$}{\lower 1.0ex\hbox{$\sim$}}}
\def\Msun {\,\mathrm{M}_\odot}
\def\Jcrit {J_\mathrm{crit}}

\def\simgreat{\lower2pt\hbox{$\buildrel {\scriptstyle >}
   \over {\scriptstyle\sim}$}}
\def\simless{\lower2pt\hbox{$\buildrel {\scriptstyle <}
   \over {\scriptstyle\sim}$}}
\def\msobh{M_\bullet^{\rm sBH}}
\def\zodot{\,{\rm Z}_\odot}
\newcommand{\lambdabar}{\mbox{\makebox[-0.5ex][l]{$\lambda$} \raisebox{0.7ex}[0pt][0pt]{--}}}

\def\na{NewA}%
\def\aj{AJ}%
\def\araa{ARA\&A}%
\def\apj{ApJ}%
\def\apjl{ApJ}%
\def\jcap{JCAP}

\def\pasa{PASA}

\def\apjs{ApJS}%
\def\ao{Appl.~Opt.}%
\def\apss{Ap\&SS}%
\def\aap{A\&A}%
\def\aapr{A\&A~Rev.}%
\def\aaps{A\&AS}%
\def\azh{AZh}%
\def\baas{BAAS}%
\def\jrasc{JRASC}%
\def\memras{MmRAS}%
\def\mnras{MNRAS}%
\def\pra{Phys.~Rev.~A}%
\def\prb{Phys.~Rev.~B}%
\def\prc{Phys.~Rev.~C}%
\def\prd{Phys.~Rev.~D}%
\def\pre{Phys.~Rev.~E}%
\def\prl{Phys.~Rev.~Lett.}%
\def\pasp{PASP}%
\def\pasj{PASJ}%
\def\qjras{QJRAS}%
\def\skytel{S\&T}%
\def\solphys{Sol.~Phys.}%

\def\sovast{Soviet~Ast.}%
\def\ssr{Space~Sci.~Rev.}%
\def\zap{ZAp}%
\def\nat{Nature}%
\def\iaucirc{IAU~Circ.}%
\def\aplett{Astrophys.~Lett.}%
\def\apspr{Astrophys.~Space~Phys.~Res.}%
\def\bain{Bull.~Astron.~Inst.~Netherlands}%
\def\fcp{Fund.~Cosmic~Phys.}%
\def\gca{Geochim.~Cosmochim.~Acta}%
\def\grl{Geophys.~Res.~Lett.}%
\def\jcp{J.~Chem.~Phys.}%
\def\jgr{J.~Geophys.~Res.}%
\def\jqsrt{J.~Quant.~Spec.~Radiat.~Transf.}%
\def\memsai{Mem.~Soc.~Astron.~Italiana}%
\def\nphysa{Nucl.~Phys.~A}%

\def\physrep{Phys.~Rep.}%
\def\physscr{Phys.~Scr}%
\def\planss{Planet.~Space~Sci.}%
\def\procspie{Proc.~SPIE}%

\newcommand{\rmp}{Rev. Mod. Phys.}
\newcommand{\ijmpd}{Int. J. Mod. Phys. D}
\newcommand{\sovjetp}{Soviet J. Exp. Theor. Phys.}
\newcommand{\jkas}{J. Korean. Ast. Soc.}
\newcommand{\PPVI}{Protostars and Planets VI}
\newcommand{\njp}{New J. Phys.}
\newcommand{\rap}{Res. Astro. Astrophys.}

\chapter[Black Hole Formation in the First Stellar Clusters]{Black Hole Formation in the First Stellar Clusters$^1$}
\label{katz}
\author[Harley Katz]{Harley Katz}
\address{University of Oxford, \\Department of Physics, Clarendon Laboratory, \\Parks Road,
Oxford OX1 3PU, Great Britain, \\ hk380@ast.cam.ac.uk }

\begin{abstract}
The early Universe was composed almost entirely of hydrogen and helium, with only trace amounts of heavy elements.  It was only after the first generation of star formation that the Universe became sufficiently polluted to produce a second generation (Population II) of stars which are similar to those in our local Universe.  Evidence of massive star cluster formation is nearly ubiquitous among the observed galaxy population and if this mode of star formation occurred at early enough epochs, the higher densities in the early Universe may have caused many of the stars in the cluster to strongly interact.  In this scenario, it may be possible to form a very massive star by repeated stellar collisions that may directly collapse into a black hole and form a supermassive black hole seed.  In this chapter, we will explore this scenario in detail to understand the dynamics which allow for this process to ensue and measure the probability for this type of seed to represent the supermassive black hole population observed at $z>6$.
\end{abstract}

\body
\section{Introduction}
\footnotetext{$^1$ Preprint~of~a~review volume chapter to be published in Latif, M., \& Schleicher, D.R.G., ''Black Hole Formation in the First Stellar Clusters'', Formation of the First Black Holes, 2018 \textcopyright Copyright World Scientific Publishing Company, https://www.worldscientific.com/worldscibooks/10.1142/10652 }

In previous chapters, we have explored the conditions for black holes to form via a direct, monolithic collapse (chapters~5 \& 6). The latter is however not the only possible pathway for black hole formation. In fact, as we saw in chapter~4, the first stars in the Universe are expected to form as clusters, and even denser and more compact clusters may form after trace amounts of metal enrichment. The first Population~II star clusters to form in the Universe may be ideal laboratories for dynamically building supermassive black hole seeds.  

In these star clusters, a very massive star (VMS) may form through repeated stellar collisions and mergers which may collapse directly to an intermediate mass black hole (IMBH) if the star becomes massive enough \citep{Heger2003}.  Because the galaxies which formed in the early Universe were significantly more dense than those we observe in the local Universe, it is very likely that the first Population~II star clusters which formed were also significantly more dense compared to those observed nearby.  Stellar densities in some of the most extreme systems in our local neighborhood can reach values of $\sim10^5$M~$_{\odot}$/yr \citep{Espinoza2009}.  At these densities, stellar interactions are expected to be frequent \citep{SPZ1999}.  The core collapse and relaxation timescales of the systems are very short compared to the age of the Universe (at the time these systems were likely to first form) as well as the main sequence lifetimes of the most massive stars\footnote{The main sequence lifetime of very massive stars is expected to plateau at $\sim3$~Myr for high mass ($\geq50$~M$_{\odot}$).} in the system.  Therefore, if these stellar densities are commonplace in the first star clusters, it is expected that the mass of the most massive star to form in the cluster is not set by the fragmentation properties of the gas, but rather by the number of mergers or collisions between high mass stars which build a VMS dynamically.

The general dynamical process for forming an IMBH seed from collisional runaway proceeds as follows:
\begin{enumerate}
\item A dense Population~II star cluster forms in a metal enriched gas cloud in the early Universe with total mass $M_c$, radius $R_c$, total number of stars $N_*$, and stellar density $n_*$.  This star cluster will have a given stellar IMF, $\xi$, virial ratio, $Q$, which determines whether the initial velocities of the star cluster are dynamically cold, hot, or virialized, a binary fraction, $b$, a fractal degree, $D$, as well as a degree of mass segregation $S$.  There is no guarantee that the initial conditions of the star cluster are spherical, but in the absence of gas, direct N-body experiments show that the star cluster will become spherical rather quickly \citep{Katz2015} and the shape of the cluster in the initial conditions only weakly affects the final outcome.
\item Once the star cluster has formed, if the IMF is not a delta function, the massive stars sink to the center of the cluster (if not already segregated at the center) due to dynamical friction.  This increases the collision/merger probability of stars in the cluster as the cross sections of these stars are much larger than their lower mass counterparts.  Once a collision or merger has occurred, the cross section of the newly formed star increases thus enhancing the probability for a future collision or merger.  This process is clearly unstable as the star with the largest cross section is most likely to dominate all subsequent collisions or mergers in the cluster.
\item After the collision/merger time scale in the cluster becomes larger than the main sequence lifetime of the VMS, the star cluster will continue to evolve normally without many more stellar collisions or mergers.  The VMS then evolves off the main sequence and depending on its mass and metallicity, it may directly collapse to an IMBH with minimal mass loss \citep{Heger2003}.
\end{enumerate}

Conceptually, this mechanism for forming a SMBH seed at high redshift is rather simple.  Furthermore, unlike two of the other scenarios which remain popular in the literature (Population III star formation and direct collapse), the conditions needed to initiate the process of collisional runaway have been observed in the local Universe -- stellar densities high enough for multiple stellar interactions have been observed in the Milky Way.  Similarly, the remnants of stellar collisions may have already been observed -- it is believed that stellar collisions may be responsible for forming blue stragglers \citep{Sanders1970,Hills1976,McNamara1976,Lombardi1996}, X-ray binaries \citep{Fabian1975}, and millisecond pulsars \citep{Lyne1987,Lyne1988}.  For this reason, collisional runaway remains worthwhile to explore the probability of this scenario being the dominant seeding mechanism for high redshift SMBHs.  

Unfortunately, the evolution of these types of star clusters is highly nonlinear and thus, properly modeling this scenario requires extremely accurate direct N-body integrators (as these systems are highly chaotic).  Furthermore, the processes of stellar evolution (single and binary, including that of a merger remnant), stellar mergers/collisions, the gas dynamics and the gravitational effects due to the larger galaxy within which the star cluster is embedded must be modeled correctly.  Finally, the initial conditions of this type of star cluster at $z>6$ are unknown.  Our current best estimates for their properties rely on detailed cosmological hydrodynamic simulations which can only resolve the scales needed to address this issue in single objects, limiting our ability to make general statements.  In what follows below and throughout this chapter, we discuss the current, state-of-the-art understanding of the collisional runaway scenario, beginning with the probability of forming a dense star cluster in the early Universe, continuing to the star cluster dynamics and stellar evolution which govern the formation of a VMS, and finishing with evaluating the probability that this scenario is the dominant mechanism for SMBH seed formation.

\section{Forming a High Redshift Birth Cloud}
Unfortunately, there is no expectation that the birth clouds in which these star clusters might form at high redshift will be observed any time in the near future.  Therefore, our entire understanding of the initial conditions of the star clusters relies on cosmological simulations and analytic modeling.  The two conditions required for the formation of a dense, bound star cluster are a metal enriched birth cloud and a high star formation efficiency.  The first criteria ensures that the birth cloud will fragment efficiently into a Population II star cluster with many stars, while Population III stars tend to form in smaller groups \citep{Bromm2002,Turk09,Stacy2010} which may prohibit collisional runaway.  The estimated metallicity for efficient fragmentation is $Z\gtrsim10^{-4}Z_{\odot}$ or $Z\gtrsim10^{-6}Z_{\odot}$ in the presence of dust \citep{Omukai2005,Schneider2012}.  The latter criterium ensures that that a bound star cluster actually forms which is a necessary requirement for collisional runaway to ensue.  The properties of a star cluster which forms in a birth cloud with an arbitrary set of properties is an unsolved problem, even at low redshift.  However, the fact that old, dense star clusters such as globular clusters exist means that at high redshift, dense birth clouds can form with a high efficiency of gas to star conversion.

\section{Metal Enriched Birth Clouds in a Cosmological Context}
\citet{Devecchi2009} used an analytic model to consider the case where metal enriched birth clouds form at the centers of atomic cooling haloes at high redshift.  In this model, they assumed that Population III star formation occurs in the low mass mini-haloes which merge to form the more massive object, therefore setting the global metallicity of the halo above the critical value needed for efficient fragmentation.  They assume that a gaseous disk forms as the baryons condense from the halo and that the Toomre parameter is sufficiently high so that the disk does not fragment efficiently.  This allows a massive star cluster to form in the center of the disk which may undergo collisional runaway.  This is put into a cosmological context by using theoretical halo mass functions as a function of redshift \citep{Press1974},  and assuming a metallicity dependent on redshift.  Finally, using a model to populate the haloes with star clusters, they check whether the system has a core-collapse time scale less than 3~Myr (i.e. the time at which the first stars are likely to explode as SN) which would be sufficient to develop a runaway instability.  The metal enrichment histories are highly uncertain so they provide multiple different models to understand the effect of this parameter \citep{Devecchi2009}. 

\citet{Katz2015} attempted a different approach where they aimed to form the birth clouds of high redshift Population II star clusters from first principles in high resolution cosmological hydrodynamical simulations.  The advantage of this approach compared to the analytic models is that one can directly measure the properties of the gaseous birth clouds from the simulation with limited additional assumptions.  The disadvantage of this method is that the physical scales required to model the dynamics of the birth cloud are small enough that this method is only practical for few numbers of objects which prohibits making general statements.  Nevertheless, the two approaches are complimentary as the former provides global statistics while the later can focus on individual properties of the cloud.

Rather than focusing on nuclear star clusters formed at the center of disk galaxies in atomic cooling haloes, \citet{Katz2015} considered the scenario of the pairwise collapse of two high redshift mini-haloes.  If the collapse of the two mini-haloes is slightly offset in time by a few Myrs, the first collapsing object will undergo Population III star formation.  The main sequence lifetimes of the average mass Population III star \citep{Hosokawa11} is expected to be short ($\sim3.9$~Myrs \citep{Schaerer2002}) and metals ejected in the supernova of these stars will enrich the gas which will fall in the second object.  If the second object is enriched to a level above the critical metallicity for efficient fragmentation, it will form Population II stars where collisional runaway may occur.

In particular, \citet{Katz2015} ran a zoom-in simulation on a pair of mini-haloes with halo masses at collapse of $M_h\sim3\times10^5$~M$_{\odot}$.  The second halo collapsed 12.9~Myr after the first (at $z=28.9$) at a distance of 117pc which is more than enough time for the first object to form Population III stars and for the metals from the supernova ejecta to reach the second halo \citep{Ritter2012}.  The evolution of the gas in the second halo was followed in great detail (at a resolution of 0.9 comoving pc $h^{-1}$ which corresponds to a physical resolution of 0.046 pc at $z=30$).  The simulations demonstrated that the mass in the center of the second mini-halo can grow to $M>10^4$~M$_{\odot}$ in less than 10~Myr within a radius of 1~pc.  The exact structure of the central cloud is subject to the resolution in the simulation but is likely to be fragmented, flattened, and ellipsoidal \citep{Katz2015}.  While the structure differs greatly depending on the simulation resolution, the mass contained within a fixed radius is very well conserved indicating that birth clouds with $M>10^4$~M$_{\odot}$ and a radius of 1~pc are robustly predicted to form in high redshift mini-haloes.  This type of birth cloud may provide the necessary conditions needed to form a dense star cluster which may undergo collisional runaway.

\section{From Birth Cloud to Star Cluster}
As described earlier, how a birth cloud with an arbitrary set of initial parameters forms a star cluster remains an open problem in modern astrophysics.  This issue is even harder to constrain at high redshift as there is very limited observational data at the metallicities and densities which are common in the early Universe.  Of particular interest are properties such as the star formation efficiency (SFE, $\epsilon$) and the stellar initial mass function (IMF).  In the local Universe, the most massive and dense cores within a molecular cloud often have embedded star clusters which form with an SFE of $\sim10-30\%$ \citep{Lada2003}.  However, \citet{Dib2011} show that the SFE in a given protostellar clump is dependent on metallicity which is primarily due to the longer time required to expel gas at lower metallicity because of weaker stellar feedback \citep{Vink2001}.  Furthermore, simulations show that compact clusters may have been subject to SFEs of $\sim60-70\%$ \citep{Pfalzner2013}.  The SFE is the most important parameter for setting the stellar mass of the star cluster.  These quantities, along with the initial stellar density are arguably the two most important parameters in determining how massive a VMS can become \citep{Katz2015}.  In order for collisional runaway to ensue, it is imperative that the cluster remains bound during its formation so that the stars will not become unbound before they can collide or merge.

Similarly, understanding the stellar IMF of the the primordial birth clouds has important implications for whether collisional runaway can proceed.  N-body simulations of star clusters show that collisional runaway is often initiated by one of the most massive stars in the cluster because it has the largest radius \citep{Katz2015}.  The rate at which these stars migrate to the center is dependent on the stellar IMF (assuming that the star cluster is not initially mass segregated).  Because collisional runaway often begins with the most massive star, the highest mass star in the IMF roughly sets the number of encounters needed to break the threshold of $\sim250~$M$_{\odot}$ needed to directly collapse to an IMBH \citep{Heger2003}.  For example, a birth cloud which forms stars up to 150~M$_{\odot}$ will have a much easier time of forming a VMS compared to an birth cloud which only created 20~M$_{\odot}$ stars as the number of collisions required to break the $250$~M$_{\odot}$ threshold increases from two to 12.

Several other parameters in addition to SFE, initial mass, initial density, and stellar IMF are needed to describe the initial state of the star cluster which forms from the primordial birth cloud.  These include density profile and structure, degree of initial mass segregation, fractal degree, and binary fraction among others.

\section{Star Cluster Evolution}
For the purposes of forming an IMBH in a primordial star cluster due to collisional runaway, there are three relevant outcomes of the star cluster evolution at low metallicity which are interesting to identify.  In the best case scenario, multiple stellar collisions lead to the formation of a VMS with $M>260~$M$_{\odot}$ which collapses to an IMBH with minimal mass loss.  A second scenario occurs when a star undergoes few collisions which creates a VMS with $260>M>150~$M$_{\odot}$ which likely results in a pair-instability supernova.  Finally, in the third scenario, not enough collisions occur and the most massive star in the cluster remains below 150~M$_{\odot}$ and the star cluster evolves normally.  The goal is to identify which parameters control the formation of a VMS and with what probability does this process occur.

\subsection{Relevant Time Scales}
Ignoring stellar evolution and the relevant hydrodynamics from any residual gas, the evolution of the star cluster can be described by entirely gravitational processes.  The dynamical state of the star cluster is dictated by the relaxation time such that the half-mass relaxation time is
\begin{equation}
t_{\rm rlx}=\left(\frac{R_c^3}{GM_c}\right)^{\frac{1}{2}}\frac{N_c}{8\ln\Lambda_c},
\end{equation}
where $R_c$ is the half-mass radius of the cluster, $G$ is the gravitational constant, $M_c$ is the cluster mass, $N_c$ is the number of stars in the cluster, and $\ln\Lambda_c\sim10$ is the Coulomb logarithm \citep{Spitzer1987}.  Because self-gravitating star clusters have negative specific heat \citep{LB1968}, they are driven to core collapse where the central density of the star cluster increases very rapidly over a finite time scale \citep{Antonov1962,Spitzer1971}.  For a system of equal mass stars, the core collapse timescale scales roughly as
\begin{equation}
t_{\rm cc}\approx15t_{\rm rlx}
\end{equation}
\citep{Cohn1980}.  The sharp increase in density in the central regions of the cluster drastically increases the probability for a stellar collision or merger.  In the case of equal mass stars, the core collapse time scale is long compared to the relaxation time scale.  

The situation is categorically different for star clusters which exhibit a range of initial masses across a stellar IMF compared to a star cluster where all stars share the same mass.  In this case, massive stars sink to the center of the cluster due to dynamical friction \citep{Chandrasekhar1943}.  The inspired time scale can be computed as 
\begin{equation}
t_{\rm df}=3.3\frac{\bar{m}}{m}t_{\rm rlx},
\end{equation}
where $\bar{m}$ is the mean mass of stars in the cluster, and $m$ is the mass of the in-spiraling star \citep{SPZ2002}.  The larger $m$ is compared to $\bar{m}$, the quicker the star will in-spiral.  Core collapse in a multi-mass cluster is then dictated by the time scale it takes for the most massive stars in the cluster to sink to the center.  If one considers a Salpeter IMF in the range $m_{\rm min}=0.1$~M$_{\odot}$ to $m_{\rm max}=100~$M$_{\odot}$, then $\bar{m}=0.39~$M$_{\odot}$ and $t_{\rm df}=0.13t_{\rm rlx}$ for the most massive star in the cluster.  Empirically, \citet{SPZ2002} find that for realistic choices of stellar IMFs, $t_{\rm cc}\approx0.2t_{\rm rlx}$.  The core collapse time scale of a multi-mass cluster is clearly much shorter than the relaxation time scale.  This suggests, that at very early times in the cluster evolution, the densities of multi-mass clusters will grow significantly in the central regions to a point where interactions are probable.   

\subsection{Collisions Versus Mergers}
A stellar merger or collision occurs when the radii of two stars overlap.  In principle, a collision may occur randomly in the densest regions of the cluster when the orbital trajectories of two stars have an impact parameter smaller than the sum of the two radii of the stars.  However, direct N-body simulations demonstrate that mergers are much more common \citep{SPZ1999}.  In this case, the massive stars at the center of the cluster form binaries dynamically which are hardened via binary exchanges or three-body interactions.  With enough interactions, the binary orbit may lose enough energy or become sufficiently eccentric such that the two stars merge.

\begin{figure}
\centerline{\includegraphics[width=10.8cm]{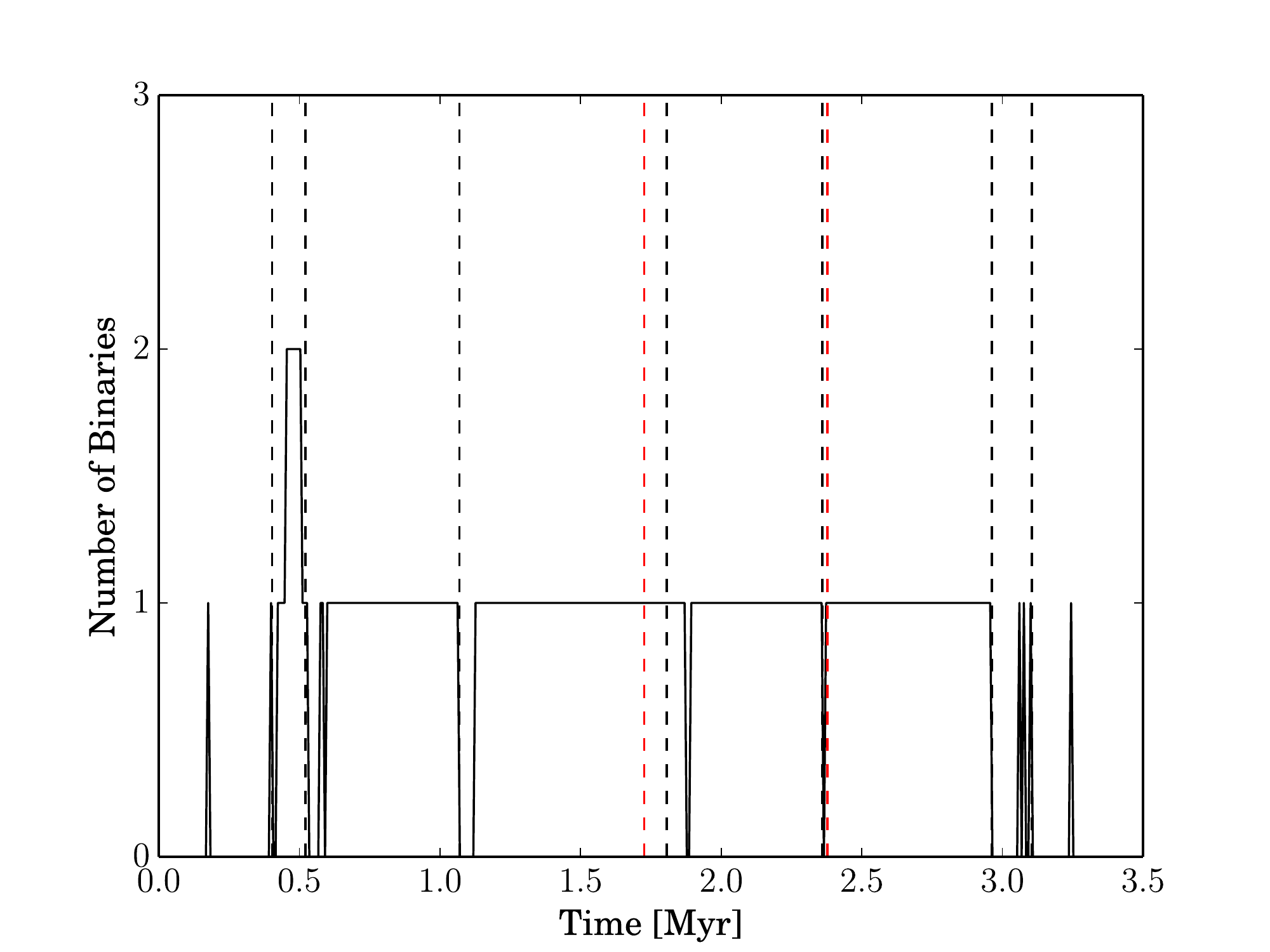}}
\centerline{\includegraphics[width=10.8cm]{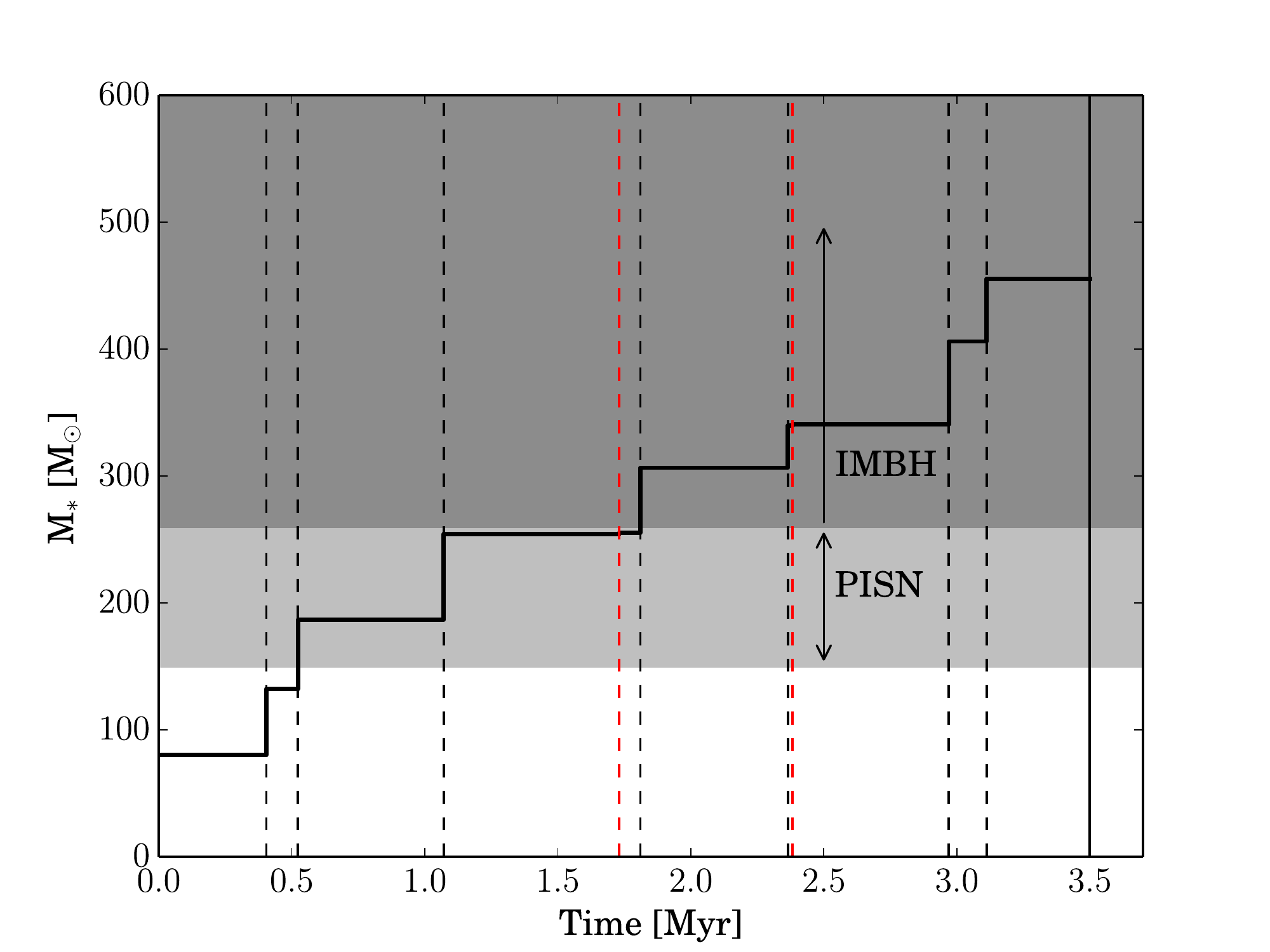}}
\caption{(Top) Number of binaries as a function of time in a star cluster simulated by \protect\citet{Katz2015}.  The dotted black vertical lines indicate the time at which a binary merger occurred in the cluster while the dotted red vertical lines indicate the time at which a stellar collision occurred.  Every time a binary merger occurs, the number of binaries in the cluster decreases by one.  (Bottom) Growth of the VMS in the same cluster as depicted in the top panel.  The stellar seed mass begins at $\sim90~$M$_{\odot}$ and grows to more than $450~$M$_{\odot}$ in less than 3.5~Myr.  The vertical black and red dashed lines are he same as in the top panel.  During a binary merger, the mass increases much more that it does via a stellar collision indicating that binary mergers are the main mechanism which dominate the growth of a VMS. Figure adopted from \citet{Katz2015}, reproduced by permission of the Oxford University Press / on behalf of the RAS.} 
\label{bincoll}
\end{figure}

In Figure~\ref{bincoll}, we show both the number of binaries as a function of time as well as the mass of the VMS as a function of time in one of the fiducial star clusters simulated by \citet{Katz2015}.  The vertical dashed black lines indicate the time at which a binary merger occurred whereas there vertical dashed red lines indicated when a stellar collision occurred.  In these simulations, the number of binary mergers which contributed to the growth of the VMS totaled seven while the number of collisions was only two.  Furthermore, from the bottom panel of this figure, it is clear that stellar collisions provide negligible amount of mass to the growth of the VMS.  Hence the growth of the VMS can be linked to the formation and hardening of binaries at the center of the cluster. 

\subsection{Growth of the VMS: Analytical Modelling}
\label{anamod}
Assuming all collisions in the cluster involve the same star, \citet{SPZ2002} provide an analytic equation for the maximum mass that a VMS can grow to such that
\begin{equation}
\frac{dm_r}{dt}=N_{\rm coll}\bar{\delta m}_{\rm coll}
\label{ana1}
\end{equation}
where $N_{\rm coll}$ is the average collision/merging rate in the cluster and $\bar{\delta m}_{\rm coll}$ is the average mass increase per collision.  Since VMS grow primarily via binary mergers, the collision/merging rate is inherently linked to the formation rate of binaries at the center of the cluster\footnote{Here we have assumed that there are no primordial binaries.}.  \citet{SPZ2002} argue that for a cluster with identical mass stars, 
\begin{equation}
N_{\rm coll}\sim f_cn_{\rm bf},
\end{equation}
where $n_{\rm bf}$ is the binary formation rate such that
\begin{equation}
n_{\rm bf}\approx10^{-3}\frac{N_c}{t_{\rm rlx}},
\end{equation}
and $f_c$ is the fraction of binaries that undergo a merger.

With the collision rate in hand, the only additional uncertainty is the average mass increase per collision.  Following \citet{SPZ2002}, this can be estimated by calculating the minimum mass star which can migrate to the center of the cluster at a time $t$ due to dynamical friction and they find that
\begin{equation}
\bar{\delta m}_{\rm coll}\approx4\frac{t_{\rm rlx}}{t}\bar{m}\log\Lambda_c.
\label{ana2}
\end{equation}

Combining Equations~\ref{ana1}-\ref{ana2}, \citet{SPZ2002} derive an analytical model for the mass at which a VMS grows for an individual cluster and find
\begin{equation}
m_r=m_{\rm seed}+4\times10^{-3}f_cM_{c0}\ln\Lambda_c\left[\ln\left(\frac{t_{\rm dis}}{t_{\rm cc}}\right)+\frac{t_{\rm cc}}{t_{\rm dis}}-1\right],
\label{mbhana}
\end{equation}
where $m_{\rm seed}$ is the mass of the star that initiates the runaway collision process, $M_{c0}$ is the initial mass of the star cluster, and $t_{\rm dis}$ is the disruption time scale of the cluster.  The disruption time scale sets the limit for how massive the VMS can grow.

Depending on context, the disruption time scale may refer to any number of effects.  In the presence of an external tidal field, $t_{\rm dis}$ is equal to the tidal disruption timescale of the cluster.  Similarly, this could also refer to the evaporation time scale of the cluster.  Although we have not yet considered how stellar evolution may affect the formation of a VMS, realistic star clusters are indeed subject to stellar evolution and it is almost certainly the main sequence lifetimes of the most massive stars in the cluster that set the disruption time scale.  While supernova feedback may not destroy or unbind the cluster, it destroys the massive stars which are the ones most likely to merge, hence preventing any more significant growth of the VMS.  For this reason, \citet{Devecchi2009} insert the main sequence lifetime of the most massive stars into Equation~\ref{mbhana} to calculate, using an analytical model, the mass of black hole seeds at high redshift.

\subsection{The Role of Stellar Evolution}
Although our previous discussion neglects the effects of stellar evolution, by no means is it unimportant.  As alluded to, for star clusters at high redshift, it is almost certainly the main sequence life time of a massive star which sets the disruption time scale rather than a dynamical process.  Primarily, the core collapse time scale must be less than the main sequence life time of the most massive stars in the cluster or else runaway stellar collisions will not occur.  Since the core collapse timescale is inversely proportional to the square root of density, this condition is almost certainly satisfied by the very high density cluster of interest at high redshift.  Runaway stellar collisions simply cannot happen in a cluster where the dynamical time scales are longer than the stellar life times.

\begin{figure}
\centerline{\includegraphics[width=10.8cm]{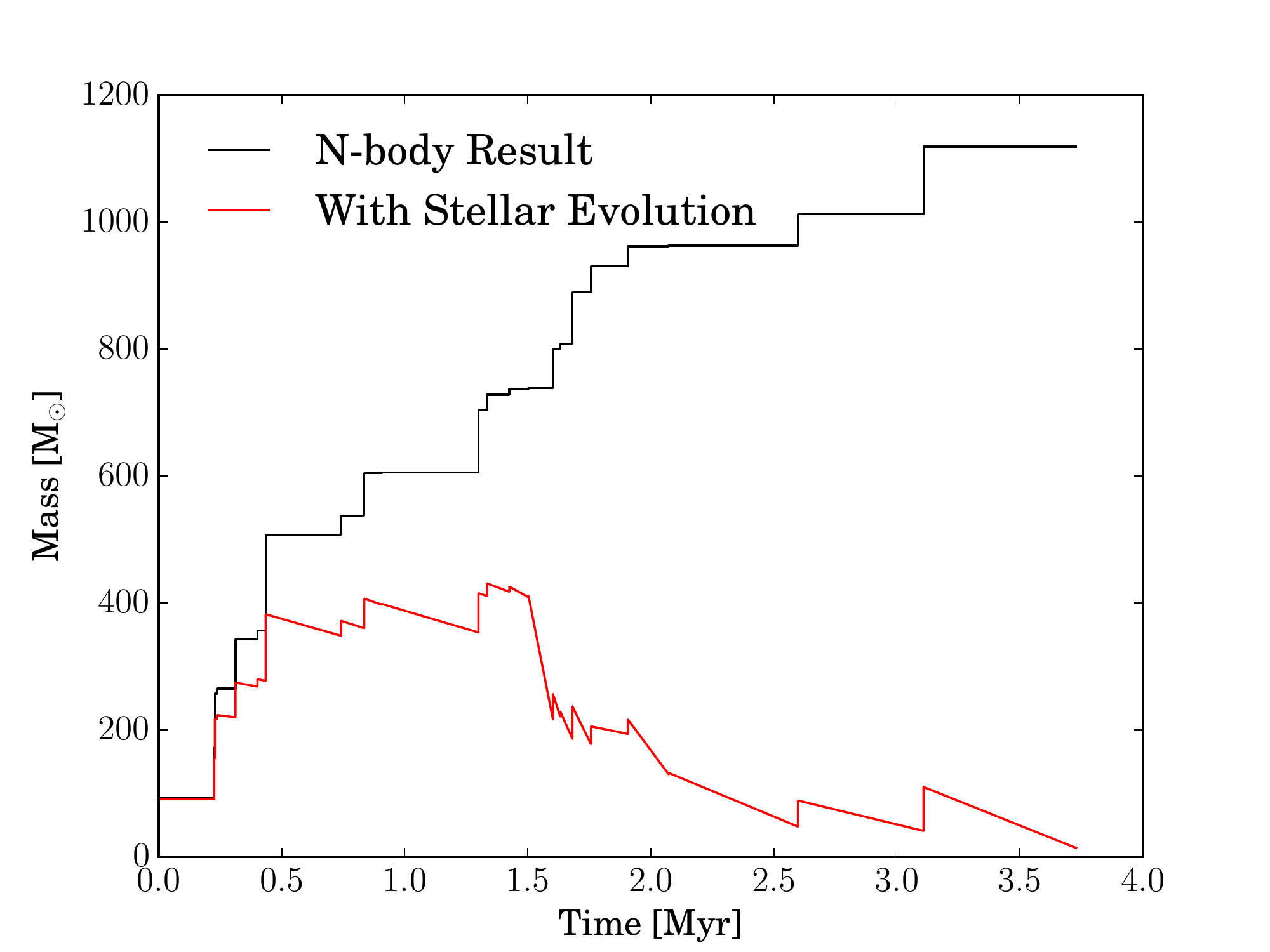}}
\caption{Comparing the growth of a VMS due to collisional runaway with an without stellar evolution at solar metallicity.  Without stellar evolution (black line) , the direct N-body experiment predicts that this star will undergo 22 collisions which grow the star from an initial mass of 92.4~M$_{\odot}$ to a final mass of 1118.9~M$_{\odot}$.  When stellar evolution, stellar winds and mass loss due to the collision are included (red line), the VMS cannot maintain a large mass and ends up having a final remnant black hole mass of 13.9~M$_{\odot}$.  The data used to make this plot was taken from \protect\citet{Glebbeek2009}. } 
\label{gwse}
\end{figure}

However, the effect of stellar winds should also not be underestimated.  The strength of stellar winds scales with metallicity so it is expected that stars at solar metallicity will undergo significantly more mass loss than those closer to primordial composition \citep{Vink1999,Vink2000,Vink2001}.  Using results from direct N-Body simulations, \citet{Glebbeek2009} calculated how massive a VMS can become if mass loss due to the collision and stellar winds are taken into account.  At solar metallicity, they demonstrate that a star which may grow to over 1000~M$_{\odot}$ due to collisional runaway in the N-Body only case may actually only become a Wolf-Rayet star of $\sim100$~M$_{\odot}$ at the end of its life due to the significant amount of mass loss when stellar evolution is considered.  In Figure~\ref{gwse}, we show the growth of a the same VMS in a simulations with and without stellar evolutionary processes from \citet{Glebbeek2009}.  While the N-body only run conserves all mass which takes part in the collision, by the end of its life time, the VMS which has formed in the run with stellar evolution has lost the majority of its mass due to stellar winds.  Even with 22 mergers, collisional runaway is very unlikely to produce a massive black hole when the cluster is at solar metallicity due to the heavy mass loss from stellar winds.  While stellar winds may prevent the formation of IMBHs in high metallicity clusters at low redshift, the gas at high redshift has a composition much closer to primordial.  \citet{Glebbeek2009} show that at lower metallicity (i.e. at $Z=0.001$ rather than $Z=0.02$) the mass loss rates are significantly lower (in this case 13 times lower) which means that the VMS can retain much of its mass.  Their simulations show that VMSs with mass $>260~$M$_{\odot}$ can form at this metallicity which suggests that even in the presence of stellar winds, collisional runaway is still a viable process for forming IMBHs at high redshift.

As was hinted at earlier, direct N-body simulations generally assume a sticky sphere approach such that if the the radii of two stars overlap, than the stars are considered to have merged.  However, smooth particle hydrodynamics simulations which have studied the mergers of stars demonstrate that not all mass which enters the collision is actually retained by the final merger remnant \citep{Dale2006,Trac2007,Glebbeek2013}.  Depending on the orientation of the collision, for example, whether it is a grazing merger or head on collision, different amounts of mass can be lost from the system into the surrounding interstellar medium \citep{Dale2006,Trac2007}.  Furthermore, the amount of mass lost in the collision also depends on the types of stars which were involved in the process \citep{Glebbeek2013}.  For low metallicity clusters, this effect may be more important than mass loss due to stellar winds \citep{Katz2015}.

In addition to the mass loss during the merger, various other physical mechanisms may affect the state of the merger remnant due to the violent process by which it formed.  Kinetic energy may be transferred into the envelope of the collision remnant which can cause the cross section to swell by a factor of $\sim100$ for some fixed time scale \citep{Dale2006}.  While this may enhance the probability of future collisions, if too much energy is provided to the envelope after it has been expanded, it can be completely driven off resulting in mass loss which would act against the formation of a VMS \citep{Dale2006}.  The disruption of the star during the merger may deliver fresh hydrogen to the core \citep{Dale2006,Glebbeek2008,Glebbeek2013}.  This can prolong the main sequence lifetime of the VMS which may allow it to grow larger due to the additional time it has to merge with other stars before undergoing supernova.  While the analytical models discussed in Section~\ref{anamod} provide important insight into the systems which produce VMSs, it is clear that these may be upper limits as stellar evolution and stellar processes can have drastic effects on the formation of a VMS.

\subsection{Growth of the VMS: N-Body Simulations}
Direct N-body simulations of sufficiently massive and dense star clusters corroborate the idea that VMSs can form in clusters due to collisional runaway with having masses $M>1000$~M$_{\odot}$ \citep{SPZ1999,SPZ2002,SPZ2004}.  Similarly without including stellar evolution, the growth of these stars is consistent with the results from analytical models \citep{SPZ2002}.  Nevertheless, these systems are already deeply nonlinear and with the inclusion of stellar processes and stellar evolution, numerical simulations are arguably the best way to evaluate the probability each cluster has in forming a VMS.  Furthermore, with advances in computing, thousands of direct N-body simulations can be run simultaneously, accelerated by GPUs, which has opened the possibility of testing various parameters in a statistically rigorous fashion \citep{Katz2015}.

\citet{Katz2015}, using star cluster initial conditions constrained from high resolution cosmological simulations, have presented one of the most complete studies which tested the multitude of star cluster parameters which may affect the growth of a VMS.  They tested the effects of varying the viral parameter, degree of initial mass segregation, the fractal degree, the initial central density, the stellar IMF, the fraction of primordial binaries, the mass of the star cluster, the initial shape of the cluster (i.e. spherical or having the same structure and rotation properties as the gas clouds in the simulation), the star formation efficiency, and the density profile of the cluster.  Furthermore, all sets of parameters were simulated multiple times to determine to probability that a star cluster with a given set of parameters forms a VMS as well as how massive the VMS becomes.

For the dense star clusters simulated in \citet{Katz2015}, core collapse occurs very rapidly compared to the main sequence life times of the most massive stars.  The star clusters are relatively low mass  (i.e. only a few$\times10^4$~M$_{\odot}$) compared to the dense globular clusters and nuclear star clusters observed in the local Universe.  In these systems they find that varying the density profile, degree of mass segregation, viral parameter and fractal degree all play minimal roles in determining how massive a VMS becomes as well as the probability of forming a VMS from the cluster.  This may be due to the fact that in this type of system, only $5-10\%$ of the cluster simulated underwent collisional runaway and formed a VMS with $M>260$~M$_{\odot}$.  Similarly, little difference was found between the spherical and non-spherical clusters.  The introduction of binaries drastically increased the number of collisions/mergers in the cluster; however, the VMSs which form in these clusters do not grow significantly larger than in clusters without as the mergers may be from binaries which do not include the VMS.  Three body interactions in the dense star cluster cause the binaries to merge and therefore, a cluster initialized with binaries is much more efficient at producing stellar mergers than without.  However, these mergers do not necessarily occur between the most massive stars nor at the center of the cluster.  Star clusters which are initialized with 100\% binaries and completely mass segregated are roughly three times less efficient at producing a VMS compared to a star cluster without primordial binaries and zero initial mass segregation.  The inclusion of binaries can effectively heat the center of the cluster and delay core collapse\citep{Heggie1975,Hut1992}.  This limits the central density of the system and thus inhibits the number of mergers which can take place.

\begin{figure}
\centerline{\includegraphics[width=6.4cm]{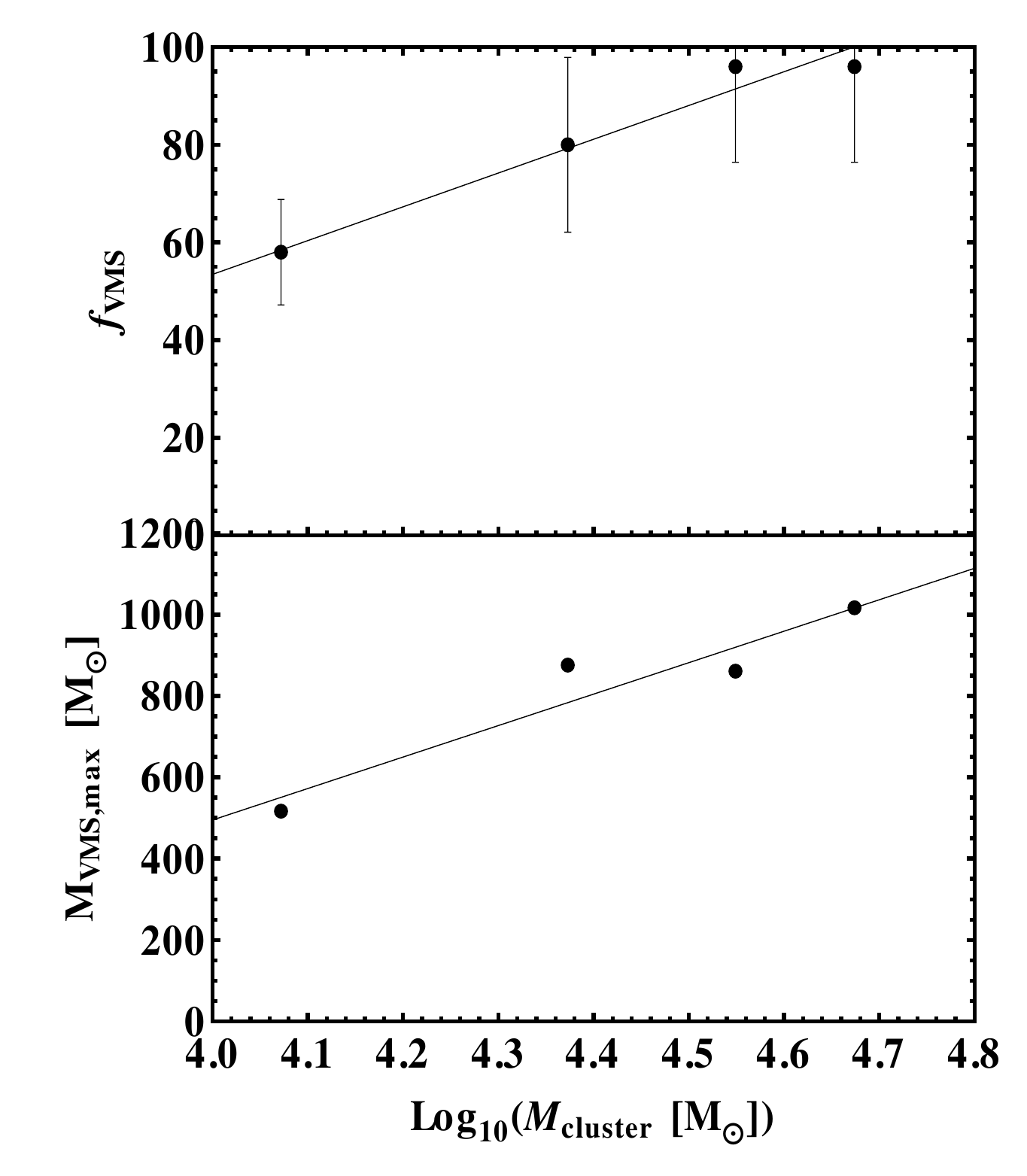}   \includegraphics[width=6.4cm]{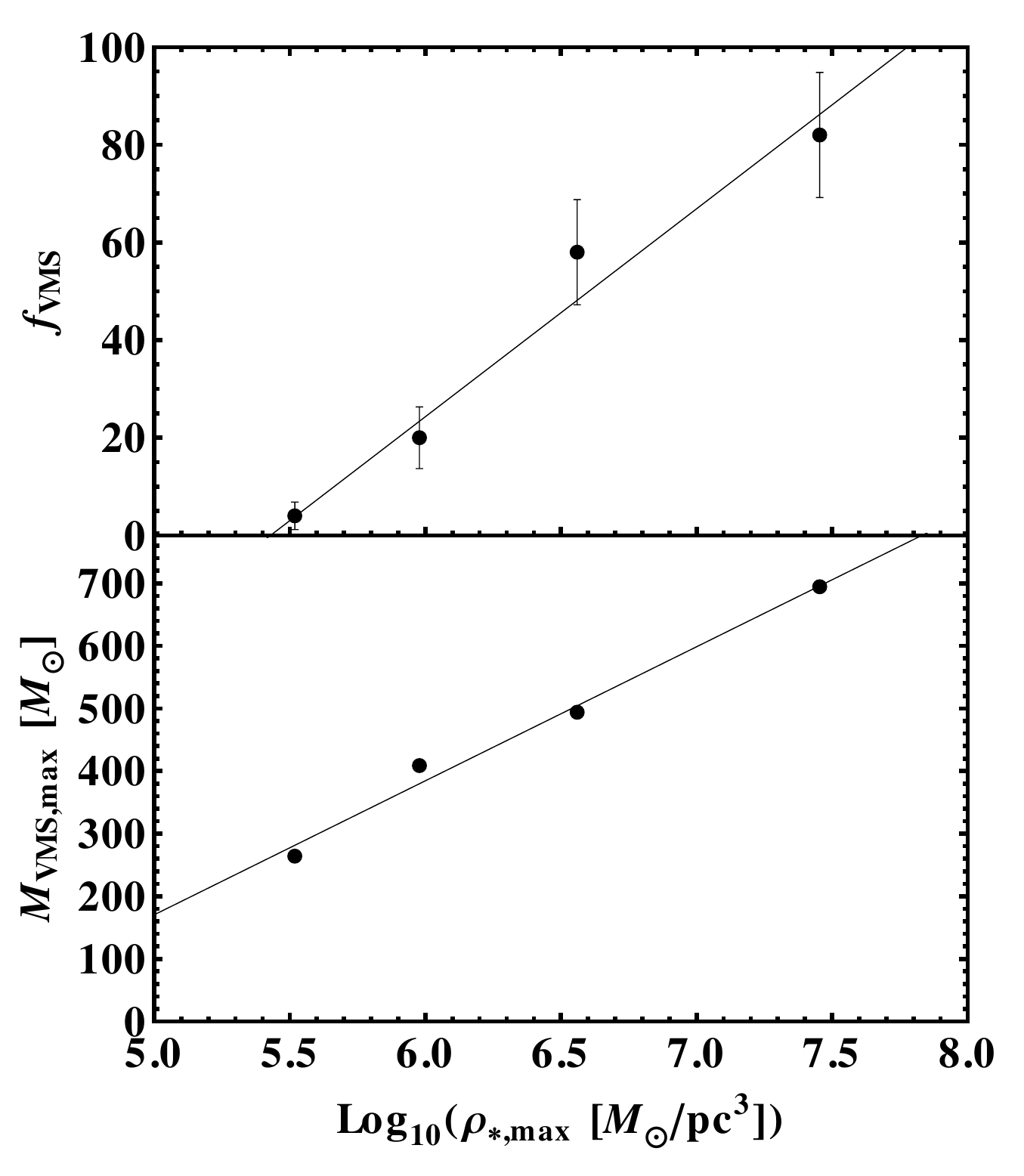}}
\caption{(Left) Probability of forming a VMS (top) and maximum mass a VMS becomes (bottom) as a function of cluster mass.  (Right) Probability of forming a VMS (top) and maximum mass a VMS becomes (bottom) as a function of initial central density.  Both the probability of forming a VMS and that mass at which a VMS grows two are extremely sensitive to the cluster mass and initial central density.  Data points represent results from simulations while solid lines represent linear fits to the data.  All data was taken from \protect\citet{Katz2015}. } 
\label{probclus}
\end{figure}

Direct N-body simulations demonstrate that the stellar IMF can have drastic differences on the mass a VMS can grow to at fixed cluster mass.  \citet{Katz2015} find that more top heavy stellar IMFs produce VMSs which are much more massive on average.  The fraction of clusters which produce a VMS is relatively consistent between the IMF but there is $\sim50\%$ difference in the maximum mass a VMS can reach between the more top heavy Flat IMFs and the more traditional Salpeter IMF.  However, most important for both the formation probability and the mass a VMS can grow to is almost certainly the initial central density of the cluster as well as the mass of the cluster.  In Figure~\ref{probclus}, we use the results from \citet{Katz2015} and show how the probability of forming a VMS in a given cluster as well as the mass of a VMS changes as a function of initial central density and mass.  As the mass and initial central density are increased, both the probability of forming a very massive star and the maximum mass a very massive star can grow to increase steeply.  In Figure~\ref{probclus}, once can see that star clusters with which have mass $M<10^5$~M$_{\odot}$ can produce VMSs with $M>1000$~M$_{\odot}$.  This mass star cluster is small compared to those observed in the local Universe, hence if these low mass systems at sufficiently high densities are indeed common in the early Universe, N-body experiments demonstrate that collisional runaway can be a promising mechanism for the formation of IMBHs at high redshift.  

\section{Predictions for Lower Redshift}
Like all other models for black hole seed formation, it is crucial that the number density of black hole seeds produced by collisional runaway is greater than or equal to the number density of observed, high-redshift, supermassive black holes.  For this model, calculating the number density of these seeds is far from straightforward as it requires knowledge of what fraction of star clusters produce a black hole and how massive these black holes become.  Numerical simulations are the best tools for determining these values and as we have shown in the previous section, the black hole seed mass is very strongly dependent on the initial mass and central density of the cluster.  Unfortunately there are still large uncertainties on the number density and properties of star clusters in the early Universe and while the cosmological hydrodynamical simulations of \citet{Katz2015} give insight into the plausibility of this process, determining the number of these seeds is much more difficult.  Assuming that the final redshift at which collisional runaway can produce a black hole is $z_{fin}=20$ and that the average mass of the black hole seed is 300~M$_{\odot}$, \citet{Katz2015} parameterized the number density of supermassive black holes as
\begin{equation}
\label{eqn:grow}
n_{\text{SMBH}}=f_pf_ff_{\text{edd}}n_{gal}(>M_{\text{thresh}}),
\end{equation}
where $f_p$ is the fraction of metal enriched haloes\footnote{We assume that the process only occurs in metal enriched halos since simulations tend to predict that Pop. III stars form either individually or in small groups \citep{Clark2008,Clark2011b,Stacy2010,Greif2011} and thus collisional runaway is unlikely to occur.} at $z_{fin}$, $f_f$ is the fraction of these haloes that form a seed black hole, $f_{\text{edd}}$ is the fraction of those seeds that can accrete constantly at the Eddington rate, and $n_{gal}(>M_{\text{thresh}})$ is the total number density of galaxies massive enough to form a star cluster in which collisional runaway can ensue.  \citet{Katz2015} concluded that for a threshold mass of $5\times10^6{\rm\ M_{\odot}}$ needed to form a high-redshift dense star cluster, if only 1\% of all star clusters formed a seed, only 1/100,000 galaxies need to be metal enriched at $z=20$ to explain the number density of observed supermassive black holes, if all seeds accrete efficiently.  There is clearly a strong tradeoff between the number of seeds which accrete efficiently and the fraction of galaxies that are metal enriched.

\begin{figure}
\centerline{\includegraphics[width=10.8cm,trim={0 0 0 10cm},clip]{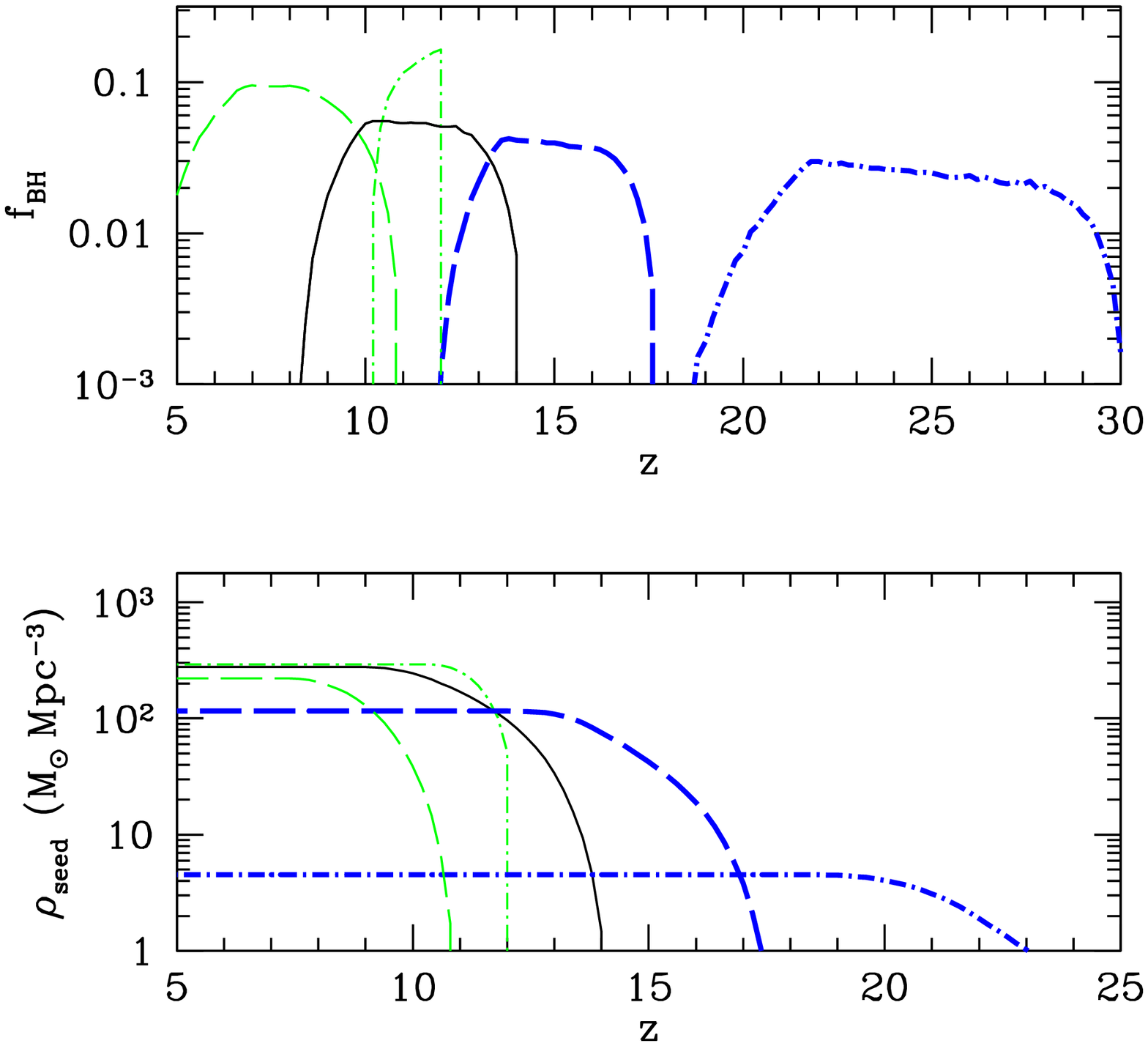}}
\caption{Mass density of supermassive black hole seeds as a function of redshift computed from the models of \protect\citet{Devecchi2009}.  The solid black line is their fiducial model which assumes that the metallicity of the system scales as $10^{-0.36z}$.  The dot-dashed blue line has a weaker scaling of $10^{-0.18z}$.  The other lines change the Pop. III star yields as well as the scatter in metallicity but the resulting mass density of black hole seeds is similar to the fiducial model at $z=6$.  In all cases, there are likely to be a sufficient number of seeds to account for the number density of supermassive black holes observed at $z=6$. Figure adopted from \citet{Devecchi2009},  \textcopyright AAS. Reproduced with permission.} 
\label{rhoofz}
\end{figure}

As described earlier, \citet{Devecchi2009} use an analytical model for the formation of high redshift primordial star clusters by assuming the form in metal enriched atomic cooling haloes.  This model is placed into a cosmological context by assuming a redshift dependent halo mass function and metal accretion history.  Knowing the mass, metallicity, and spin of each halo allows them to calculate the star cluster parameters as a function of redshift and determine which systems will undergo collisional runaway.  Thus, number densities of black hole seeds can be computed as a function of redshift.  In Figure~\ref{rhoofz}, we show the comoving mass density of supermassive black hole seeds as computed by \citet{Devecchi2009} for five different models where the metallicity-redshift relation is varies, the scatter in this relation is varied, and the critical metallicity needed to form a Pop~II star cluster is also varied.  The variations in these models produces a range in black hole seed mass densities from $\sim4{\rm\ M_{\odot}/Mpc^3}-300{\rm M_{\odot}/Mpc^3}$ by $z=6$.  The seed masses in their model are typically $\sim1000-2000{\rm\ M_{\odot}}$ and thus dividing the mass density by the mean seed mass give an average number density of $\sim2.6{\rm Gpc^{-3}}-200{\rm Gpc^{-3}}$.  Note that in all cases, the model of \citet{Devecchi2009} seems to be above the observed number density of high redshift supermassive black holes at $z\sim6-7$ which has a lower limit constraint of $1.1{\rm Gpc^{-3}}$ \citep{venemens13}. 

\section{Conclusion}
We have described here both the physics of high redshift star clusters that lead to runaway stellar conditions as well as the formation mechanisms for these types of objects in the early Universe.  The key conditions are that the core collapse time scale of the star cluster is significantly less than the lifetimes of the main sequence stars.  This is a necessary, but insufficient condition that leads to collisional runaway.  The probability that a star cluster undergoes this process is highly dependent on the initial central mass and density of the system \citep{Katz2015}.

Although the number density of these systems is highly uncertain, both analytical calculations and numerical simulations seem to suggest that it is plausible that a sufficient number of these IMBHs can be produced to explain the number density of $z=6$ supermassive black holes \citep{Katz2015,Devecchi2009}.  The major open questions are whether or not these systems can accrete efficiently enough to grow supermassive within the first billion years.  Furthermore, the properties of high redshift star clusters are still highly unconstrained.  The latter is being addressed by state of the art numerical simulations that are beginning to sample multiple haloes and constrain the properties of primordial star clusters \citep{Sakurai2017}.  The former is a particularly difficult problem numerically however, some analytical work suggests that for star clusters which strong gas inflows, certain conditions may lead to super-Eddington growth \citep{Alexander2014}.  Likewise, mergers with other stellar mass black holes may allow the dominant object to accrete mass extremely efficiently \citep{Davies2011ApJ,Lupi2014}. Very recent studies now investigated collisions also in Population III clusters \citep{Reinoso18} and explored the interaction between collisions and accretion \citep{Boekholt18}. These topics certainly deserves further exploration and provide a promising channel for the formation of supermassive black hole seeds.

While we already discussed the statistical predictions of the collision-based scenario and the comparison with observations, a detailed account on the current observational status is given in chapter~12, and predictions of other scenarios are outlined in chapter~9. Future observational probes of the masses of the first black holes are outlined in chapter~14. The next chapter~8 will describe the evolution of supermassive stars in more detail. Subsequent accretion physics, in particular the super-Eddington case, will be discussed in chapters~11 and 12.

{
\bibliographystyle{ws-rv-har}    
\bibliography{ref}
}

\printindex[aindx]           
\printindex                  

\end{document}